# Measurement of Gravitomagnetic and Acceleration Fields Around Rotating Superconductors

Martin Tajmar, Florin Plesescu, Bernhard Seifert, Klaus Marhold

*Space Propulsion, ARC Seibersdorf research GmbH, A-2444 Seibersdorf, Austria*
*+43-50550-3142, martin.tajmar@arcs.ac.at*

**Abstract.** It is well known that a rotating superconductor produces a magnetic field proportional to its angular velocity. The authors conjectured earlier, that in addition to this so-called London moment, also a large gravitomagnetic field should appear to explain an apparent mass increase of Niobium Cooper-pairs. A similar field is predicted from Einstein's general relativity theory and the presently observed amount of dark energy in the universe. An experimental facility was designed and built to measure small acceleration fields as well as gravitomagnetic fields in the vicinity of a fast rotating and accelerating superconductor in order to detect this so-called gravitomagnetic London moment. This paper summarizes the efforts and results that have been obtained so far. Measurements with Niobium superconductors indeed show first signs which appear to be within a factor of 2 of our theoretical prediction. Possible error sources as well as the experimental difficulties are reviewed and discussed. If the gravitomagnetic London moment indeed exists, acceleration fields could be produced in a laboratory environment.

## INTRODUCTION

Since the measurements of WMAP (Spergel et al., 2003), we know that the universe is best described as being flat and it has a non-zero cosmological constant Λ. A well known cosmological consequence is the observed acceleration of the expansion of the universe. Under these circumstances, gravity can be very well described by a first-order approximation or linearization of Einstein's general relativity theory. Nearly all assessments up to now only considered standard general relativity with no cosmological constant. This leads to the so-called Einstein-Maxwell equations, splitting gravity into gravitoelectric and gravitomagnetic (or frame-dragging) fields very similar to electromagnetism (Forward, 1961; Tajmar and de Matos, 2001). This has therefore been dubbed gravitoelectromagnetism. However, measurements such as WMAP tell us that this is not the complete description as we are missing the influence of the cosmological constant. Linearizing Einstein's equation including Λ leads to a set of equations which are very similar to the Proca equations thst describe massive spin-1 fields (e.g. including the effect of a massive photon in electromagnetism) (Tajmar, 2006; Argyris and Ciubotariu, 1997; de Matos, 2006). The presence of a cosmological constant can be therefore interpreted as a consequence of a massive graviton (or at least a "spin-1" like graviton).

Recent work by one of the authors showed that in order to observe a constant dark energy density throughout the universe as measured by WMAP, the graviton mass must depend on the local density of matter (Tajmar, 2006). This would enable us to investigate the effect of a cosmological constant not only on the scale of the universe but also in a laboratory environment. The effect of a non-zero photon mass is illustrated by a superconductor. According to quantum field theory, gauge symmetry breaking in a superconductor leads to a massive photon where its wavelength is observed as the penetration depth. Using Proca equations and the massive photon, it is straightforward to show two basic effects that characterize a superconductor: shielding of electromagnetic fields inside the superconductor (Meissner-Ochsenfeld effect) and the generation of a magnetic field due to a rotating superconductor (London moment) (de Matos and Tajmar, 2005). Note that the London moment is generated in addition to the usual magnetic field in standard Maxwell theory such as the one resulting from a flowing current. One quickly realizes the difference as the London moment does not depend on the magnetic permeability of vacuum $\mu_0$ which is usually common to all classical magnetic phenomenons. Although the field originates from the massive photon which is present only in the interior of the superconductor, the London moment is also measured outside of the superconductor due to the quantization of the canonical moment.

Using Proca-Einstein equations, one arrives at similar conclusions for the case of gravitational fields. The presence of a non-zero spin-1 graviton mass leads to a gravitomagnetic London moment (de Matos and Tajmar, 2005, Tajmar and de Matos, 2006a). Using our dark-energy compatible graviton mass solution, the gravitomagnetic London moment is $B_g=-2\omega$ for classical matter. This is nothing else then the so-called gravitomagnetic Larmor Theorem (Mashhoon, 1993), describing the inertial properties of matter in a rotating reference frame. This can be considered as the foundation of classical mechanics. However, as this field is defined by the local density of matter, it is only present inside the rotating material, similar to all inertial or pseudo-forces. In case of a superconductor, this should change as also the "classical" London moment is measured outside the superconductor due to the quantization of the canonical moment which now also includes a non-negligible gravitomagnetic component. Since the graviton mass shall depend on the density of matter but only the contributions from the Cooper-pairs count for the quantized canonical momentum, we should be able to measure a gravitomagnetic field outside the superconductor given by:

$$B_g = 2\omega \cdot \frac{\rho^*}{\rho} \quad , \tag{1}$$

where $\rho^*$ is the Cooper-pair mass density and $\rho$ is the bulk density of the superconductor (the sign is reversed due to Becker's argument that the Cooper pairs are lagging behind the lattice (Becker, Heller and Sauter, 1933)). For Niobium we get $B_g=3.9\times10^{-6}\cdot\omega$ rad.s$^{-1}$ at T=0 K. This is nearly two orders of magnitude below a recently published conjecture by one of the authors, that a non-classical gravitomagnetic field could be responsible for a reported disagreement between the theoretically predicted and measured Cooper-pair mass of a rotating Niobium ring (Tajmar and de Matos, 2003; 2005; 2006b). Although the values for the gravitomagnetic London moment are predicted to be small, compared to classical gravitomagnetic fields, e.g. generated by the rotating Earth, they are enormous.

In order to test our theory, an experimental program was established at ARC Seibersdorf research. We concentrate on two aspects: (i) a rotating superconductor is predicted to generate a large gravitomagnetic field. This field can be directly measured using gyroscopes, similar to the approach of Gravity-Probe B. (ii) A time-varying gravitomagnetic field should induce a gravitoelectric field similar to the Faraday induction law. This can be detected using very sensitive accelerometers. If we consider a superconducting ring which is angularly accelerated, a gravitational field should be induced opposing its cause (gravitational Lenz law) following the induction law,

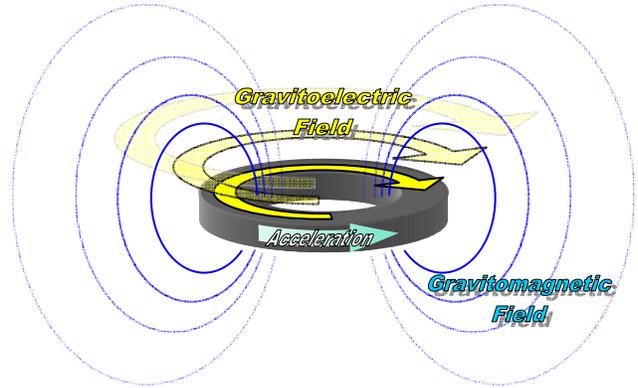

**FIGURE 1.** Gravitomagnetic and Gravitoelectric Field Generated by a Rotating and Angularly Accelerated Superconductor.

$$g = -\dot{B}_g \frac{r}{2}\cdot\hat{\varphi} = -\frac{\rho^*}{\rho}r\dot{\omega}\cdot\hat{\varphi} \quad , \tag{2}$$

where $r$ is the radial distance inside the ring and $\hat{\varphi}$ the azimuthal unity vector. A short illustration is shown in Fig. 1. The following paper will describe our experimental apparatus and the results obtained so far.

# EXPERIMENTAL FACILITY

The core of our setup is a rotating superconducting ring with an outer diameter of 150 mm, a wall thickness of 6 mm and a height of 15 mm inside a large cryostat. The ring can be rotated using a brushless servo motor or a pneumatic air motor to minimize any electromagnetic influence. Angular velocities up to 650 rad.s$^{-1}$ and a peak angular acceleration of around 1500 rad.s$^{-2}$ could be achieved. From these numbers, the requirements for our sensors can be derived using Equs. (1) and (2). In order to obtain a signal-to-noise ratio of at least 10:1 for a Niobium superconductor, we must be able to have a resolution of 2x10$^{-4}$ rad.s$^{-1}$ on the gyroscope to detect the gravitomagnetic field inside the superconducting ring. In our setup, the sensors are placed above the ring; therefore an even higher resolution of 2x10$^{-5}$ rad.s$^{-1}$ is necessary as the field is expected to decrease similar to electromagnetic fields over distance. The accelerometer on the other hand needs a resolution of at least 1x10$^{-5}$ m.s$^{-2}$ or 1 micro-g (in units of the standard Earth acceleration) for the same signal-to-noise ratio of 10:1 if the sensor is placed inside the ring at a radial distance of 3.6 cm. Niobium seems to be a good starting choice as its $\rho^*/\rho$ ratio is high compared to other superconductors and its critical temperature of 9.3 K is about 5 K above the liquid helium temperature (Tajmar and de Matos, 2006a). High-Temperature superconductors such as YBCO have a $\rho^*/\rho$ ratio which is more than an order of magnitude below Niobium which would drive the sensor requirements also more than one order of magnitude down.

**TABLE 1.** Overview of Available Gyroscopes.

| Gyroscope Technique | Products | Noise Level (rad.s$^{-1}$.Hz$^{-0.5}$) | Dimension (mm) | Price (k$) |
|---|---|---|---|---|
| Ring Laser Gyro | Honeywell GG1320AN | 1x10$^{-6}$ | 87 ⌀, 45 | 27 |
| Fibre Optic Gyro | KVH DSP-3000 | 2x10$^{-5}$ | 89 x 48 x 33 | 4 |
| Solid State Vibrating Structure | BAE Systems VSG | 3x10$^{-3}$ | 40 x 43 x 58 | 1 |
| MEMS Gyro | O-Navi Gyropak 3 | 1x10$^{-3}$ | 20 x 20 x 4 | 0.1 |

From Table 1 it is clear that only laser/fibre optic gyros fulfill our resolution requirement of 2x10$^{-5}$ rad.s$^{-1}$. A very good property of gyros is the fact that they are very insensitive to the electromagnetic or vibration environment. For example, the Honeywell GG1320AN has a magnetic offset of 10$^{-4}$ rad.T$^{-1}$. The maximum magnetic field from the electric motor at the sensor location was found to be 50 μT which leads to an offset of 10$^{-9}$ rad that is well below the gyro's resolution. We selected the KVH DSP-3000 gyro for the measurements as it promised a good signal-to-noise ratio at moderate costs.

**TABLE 2.** Overview of Available Accelerometers.

| Products | Noise Level (μg.Hz$^{-0.5}$) | Dimension (mm) | Price (k$) |
|---|---|---|---|
| Honeywell QA3000 Q-Flex | 0.05 | 25 x 25 x 25 mm | 5.5 |
| Colibrys Si-Flex SF1500S | 0.3 | 25 x 25 x 25 mm | 0.6 |
| Silicon designs 1221L-02 | 2 | 8 x 8 x 3 mm | 0.2 |
| MWS Sensorik BS 5401 | 25 | 23 x 18 x 16 mm | 2.3 |

For the accelerometers, it turned out that background noise during the experiments was on the order of a couple of μg, therefore the Honeywell and Colibrys sensors seemed appropriate. We also required an accelerometer where the offset due to magnetic fields was small. We selected the Colibrys Si-Flex SF1500S, which is a capacitive sensor measuring the tilt between two silicon plates which are unaffected by magnetic fields (e.g. the Silicon designs 1221L-02 uses Nickel plates where the offset due to the magnetic field from the electric motor was orders of magnitude above the gravitomagnetic London moment prediction).

The experimental setup is illustrated in Fig. 2. The motor and the superconductor assembly are mounted on top and inside a liquid helium cryostat respectively, which is stabilized in a 1.5 t box of sand to damp mechanical vibrations induced from the rotating superconductor. The accelerometers and gyros are mounted inside an evacuated chamber made out of stainless steel which acts as a Faraday cage and is directly connected by three solid shafts to a large structure made out of steel that is fixed to the floor and the ceiling. The sensors inside this chamber are thermally isolated from the cryogenic environment due to the evacuation of the sensor chamber and additional MLI isolation covering the inside chamber walls. Only flexible tubes along the shafts (necessary to seal the cryostat) and electric wires from the sensor chamber to the upper flange establish a mechanical link between the sensor chamber and the cryostat. This system enables a very good mechanical de-coupling of the cryostat with the rotating superconductor and the sensors even at high rotational speeds. In order to obtain a reliable temperature measurement, a calibrated silicon diode (DT-670B-SD from Lakeshore) was installed directly inside the superconductor. A miniature collector ring on top of the motor shaft enabled the correct readout even during high speed rotation. Two temperature fixpoints enabled a temperature calibration during each run: the liquid helium temperature of 4.2 K and the evaluation of the critical temperature of the superconductor using the field coil. When the superconductor was cooled down, the field coil was switched on with a field of about 300 µT, orders of magnitude below the critical field strength from the superconductors used. A Honeywell SS495A1 solid state Hall-sensor was installed inside the sensor chamber. Initially, the superconductor acted as a magnetic shield. But when the superconductor passed $T_c$, the magnetic field from the coil was recorded on the Hall sensor. At this point in time, the temperature read-out from the silicon diode must then correspond to the critical temperature.

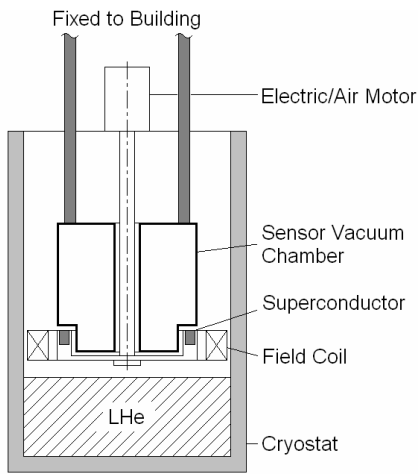  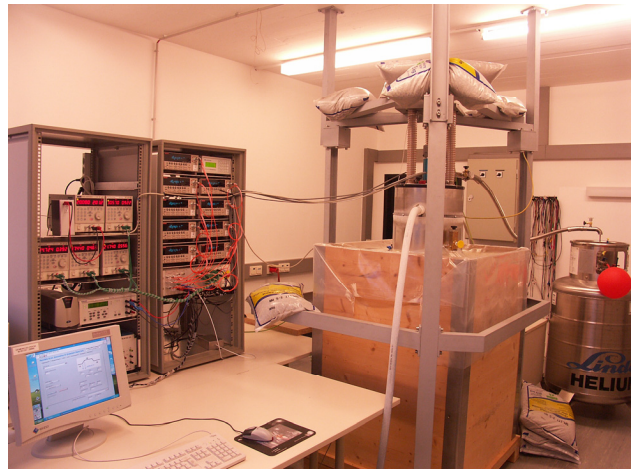

**(a)** Schematic Setup.    **(b)** Facility in the Lab at ARC Seibersdorf research.

**FIGURE 2.** Experimental Setup.

The sensor-chamber can be equipped with accelerometers or gyros. The different configurations are shown in Fig. 3. Initially, the accelerometer measurements were carried out with tangential, radial and vertical sensors (single-configuration) on three levels: In-Ring (maximum field expected), Above-Ring (smaller field expected due to field expansion) and Reference (mechanical environment, at least two orders of magnitude less signal that in-ring). All sensors are mounted on the same mechanical structure. By observing the difference between e.g. the in-ring and the reference position, mechanical artifacts such as tilts are compensated and only fields emitted from the superconductor should be measured. The radial and axial distances for each sensor are shown in Fig. 4a.

At a later stage, the accelerometer measurements focused only on the tangential direction, where the effect from the gravitomagnetic London moment is expected. Therefore, a so-called Curl-Configuration was implemented where 4 tangential sensors were equally distributed on the three levels (according to the Einstein-Proca/Faraday induction law, curl gravitoelectric fields are expected similar to the electric fields in a transformer). Each accelerometer has a mirror partner where the opposite sign of the signal is expected. By subtracting one signal from its mirror partner

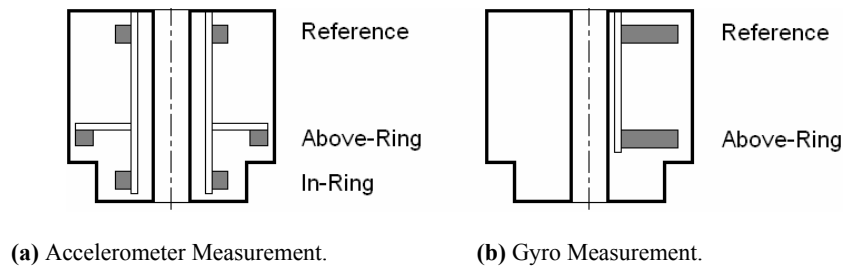

**(a)** Accelerometer Measurement.    **(b)** Gyro Measurement.

**FIGURE 3.** Sensor Chamber Confuguration.

signal, only a curl field remains and all sensor related offsets (e.g. from vibration or temperate changes) are eliminated. The accelerometers were read out using Keithley 2182 Nanovoltmeters with a measurement rate of 10 Hz. In order to measure all acceleration positions at the same time, each sensor was connected to its own nanovoltmeter.

For the gyros, only two positions were filled (reference and above-ring) due to the larger size of the sensors (see Fig. 3b). Their radial distance is 5.4 cm and the axial distances are 4.3 cm (above-ring) and 21.6 cm (reference) measured from the top of the superconductor. They have a digital output and can be directly connected to the computer for data processing.

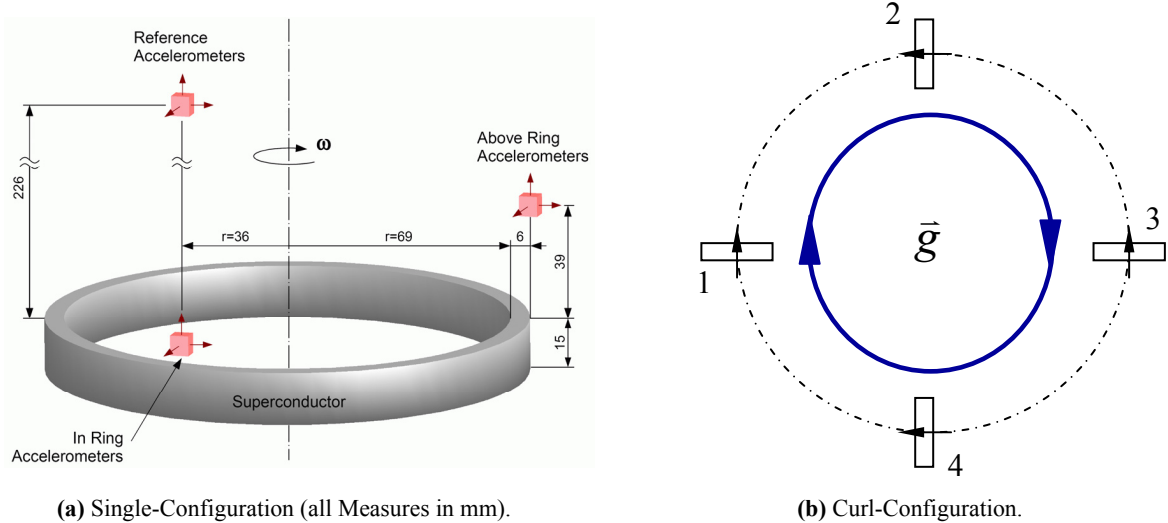

**(a)** Single-Configuration (all Measures in mm).      **(b)** Curl-Configuration.

**FIGURE 4.** Position of Accelerometers.

## ACCELEROMETER MEASUREMENTS

The following chapters will describe the external sensor influence and accelerometer measurements in the single- and the curl-configuration around angularly accelerated superconductors.

### Evaluation of Sensor External Influence and Overall Performance

In more than three years various facility updates were implemented to reduce the external mechanical, thermal and electromagnetic influence on the accelerometers. At the end, the sensors using differential readout in vertical position had a noise level with a sigma of 5 µg and the radial and tangential sensors of about 15 µg. The most important external influences were as follows:

− As soon as the low-temperature bearings degraded or if the rotating axis got in resonance (usually at an angular velocity of about 400 rad.s$^{-1}$), the resulting acoustic noise triggered a negative offset on all sensors. A similar effect was also observed during helium evaporation when the superconductor started to rotate in the liquid. This offset was further evaluated by mounting a single accelerometer on a shaker table where the sensor output could be observed during various external accelerations at different frequencies (see Fig. 5). This actually triggered the need for the "Curl-Configuration" to cut out all sensor-related offsets from the data which was implemented at a later stage. Significant effort was made to damp all noise sources inside the cryostat. However, as the noise level is independent of the sense of rotation, the offsets can be clearly identified by always performing two slope measurements after each other with alternating sense of rotation.

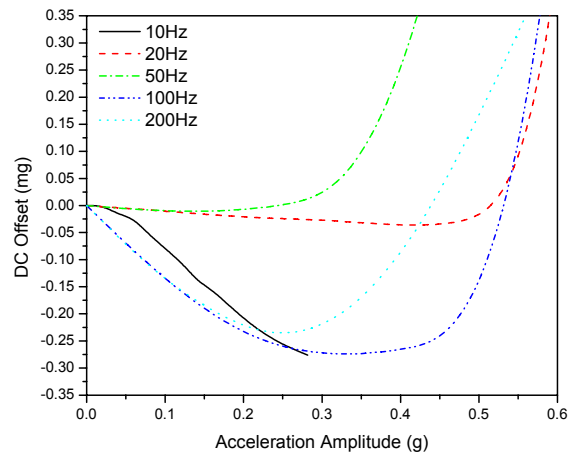

**FIGURE 5.** Accelerometer Offset Due to Vibration.

- Using the field coil, the influence of a strong magnetic field on the accelerometers was evaluated. The test was done with a BSCCO superconductor at 117 K (normal conductive) and 77 K (superconductive). An oscillating magnetic field with an amplitude of 20 mT was applied and the sensor responses were evaluated. The sensor offsets were found to be linearly proportional to the applied field, at 20 mT a maximum offset of 10 µg was measured. As the maximum magnetic field from the electric motor was measured to be about 50 µT, the maximum offset for the real runs should be therefore less than 0.025 µg, which is far below the measurement threshold of the sensors (about 1 µg). As the London magnetic field developed by the superconductor is in the nT range at full speed, the influence of magnetic fields on the results can be completely neglected.

## Single Sensor Configuration

First, high temperature superconductors were tested using liquid nitrogen. Rings of BSCCO and YBCO (slightly larger outer diameter of 160 mm and wall thickness of 15 mm for better stability of the ceramic) were rotated with maximum accelerations at room temperature and at 77 K. All sensors (tangential, radial and vertical on all levels) showed no significant signal difference between both temperatures within the measurement resolution. This puts a limit on the coupling factor between induced gravitational fields and applied angular acceleration of $\pm 1 \times 10^{-8}$ g.rad$^{-1}$.s$^2$. According to Equ. (2), we would have expected a $g/\dot{\omega}$ factor for BSCCO and YBCO of around $-5 \times 10^{-10}$ g.rad$^{-1}$.s$^2$, which is more than one order of magnitude below. Therefore, this negative result was to be expected. We also used YBCO under liquid helium conditions down to 4.2 K obtaining similar results.

Next, Niobium was tested under LHe conditions. For the first time, the tangential accelerometers showed signals that reacted to the applied acceleration on the superconducting ring. Example of raw data is shown in Fig. 6, where both the tangential sensor data (differential meaning that the mechanical reference position was subtracted) as well as the applied tangential acceleration to the superconductor is shown. The temperature during these runs varied between 4 and 6 K well below the superconductor's critical temperature of 9.3 K and the sampling frequency was 10 Hz. As the rotating shaft has a resonance frequency of about 400 rad.s$^{-1}$, the sensor signals were damped by a factor of 5 for angular velocities above 350 rad.s$^{-1}$ to reduce the mechanical vibration noise from the expanding helium and the low temperature bearing (general procedure also for all other accelerometer measurements). All applied acceleration profiles shown were obtained when the angular velocity was below 350 rad.s$^{-1}$, which means that the measurements are not influenced by this damping. Note that the sign of the peaks change together with the sign of the applied angular acceleration. This rules out vibration induced offsets as discussed in the external sensor influence section above. Significantly lower peaks were observed immediately when the superconductor passed $T_c$. A few Kelvin above $T_c$ no peaks were observed any more in all data. The peaks were found out to be proportional to the applied acceleration (Tajmar et al, 2006) as expected from Equ. (2).

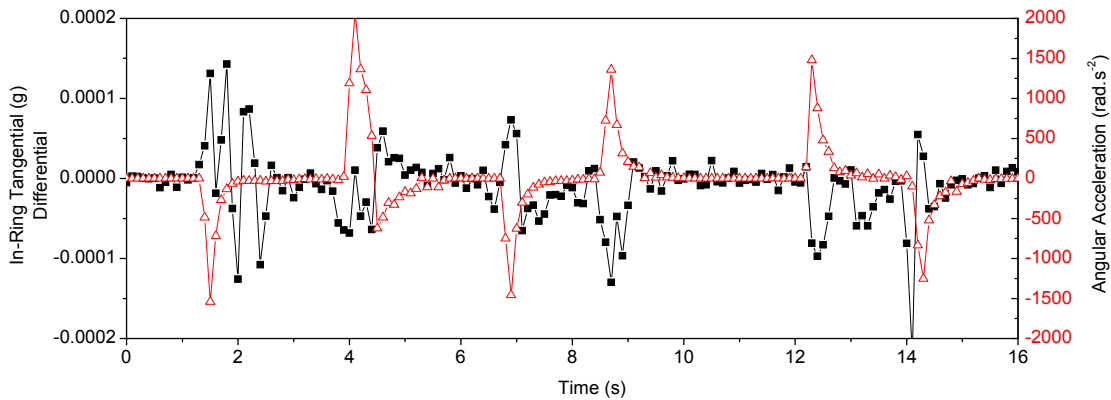

**FIGURE 6.** Example of Raw Data for In-Ring Tangential (Differential) Sensor Data Versus Applied Angular Acceleration at T=4-6 K (■ … Accelerometer Signal, ∆ … Applied Angular Acceleration to Superconductor).

Fig. 7 shows a signal averaged plot consisting of 4 profiles which were recorded at a temperature of 5 K. At an angular acceleration of 1200 rad.s$^{-1}$, a reaction of the accelerometers on the order of 35-75 µg was recorded. The coupling factor is -3.9±0.52×10$^{-8}$ g.rad$^{-1}$.s$^2$. That is about a factor of 5 above the theoretical prediction in Equ. (2).

Is this accelerometer signal indeed caused by induced gravitomagnetic fields, and why are they larger than the theoretical prediction? The accelerometer reacts to all accelerations such as tilts and not only to gravitational fields. Moreover, the accelerometer signal oscillates following its operational principles (the two silicon plates act as an oscillator). This could have caused the enhancement of the peaks in the data leading to an overprediction of the effect. In the next section we will describe measurements using the curl-configuration which should cut out not only sensor offsets but also any mechanical artifacts (tilts and torques on the mechanical structure where the sensors are mounted). Fig. 6 and 7 were performed with the air motor to eliminate any electromagnetic influence.

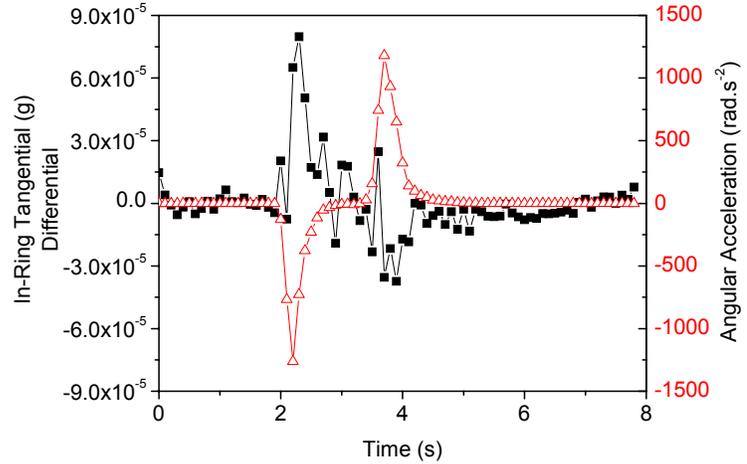

**FIGURE 7.** Signal Averaged over 4 Measurements at T=5 K; In-Ring Tangential (Differential) Sensor Data (■) Versus Applied Angular Acceleration (Δ).

## Curl Sensor Configuration

The following runs were carried out with only tangential accelerometers where each sensor has a mirror partner as described in Fig. 4 with a Niobium superconductor. By subtracting e.g. Sensor 1 – 3 (its mirror partner), then only a curl-field is measured as predicted from Equ. (2). Fig. 8 shows this example with signal averaged plots from selected profiles for the curl acceleration field and the applied angular acceleration. The difference between superconducting ($g/\dot{\omega}$ = -2.26±0.3×10$^{-8}$ g.rad$^{-1}$.s$^2$) and normal conducting ($g/\dot{\omega}$ = -1.24±1×10$^{-9}$ g.rad$^{-1}$.s$^2$) is clearly visible. Also the correlation between measured acceleration and applied acceleration is good (0.78), only the second sensor peak precedes the applied acceleration for 0.2 s. That could probably relate to the increase in temperature from 4.5 to 6.5 K during the profile which also affects the Cooper-pair density.

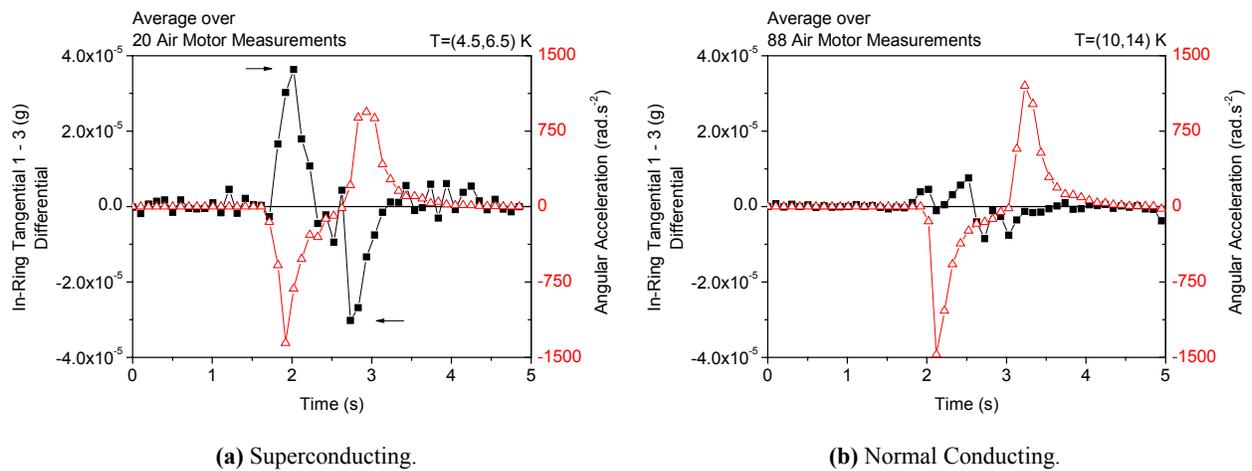

**(a)** Superconducting.   **(b)** Normal Conducting.

**FIGURE 8.** Signal Averaged In-Ring Tangential 1 – 3 (Differential) Sensor Data (■) Versus Applied Angular Acceleration (Δ).

In this curl configuration, the mechanical artifacts are effectively reduced and a signal-to-noise ratio of about 15:1 could be achieved. This greatly adds confidence that the effect as it is observed is real – whatever cause it has. Also the similarity between the coupling factor in curl- and single-sensor configuration (Fig. 7 and Fig. 8a) is very encouraging.

Fig. 9 shows the in-ring coupling factor from more than 100 profiles signal averaged in temperature intervals of one Kelvin. The red curve overlays Equ. (2), our theoretical prediction based on the ratio of Cooper-pair mass- and bulk density, multiplied by a factor of 2 which gives a reasonable fit. The results indeed look like if the observed effect relates to the Cooper pair density as predicted. The data points at 10 and 12 K fall off the prediction, nevertheless the trend following Equ. (2) is clearly seen. Further reduction of mechanical noise (e.g. pump on the evaporated helium to reduce the pressure in the cryostat) and better sensor resolution (e.g. using the Honeywell QA3000) could hopefully lead to clearer results. Following Fig. 9, the coupling factor (observed tangential acceleration versus applied angular acceleration) at T=0 K is $-14.4 \pm 2.8 \times 10^{-9}$ g.rad$^{-1}$.s$^2$. This data only represents a curl-configuration and corrects the numbers from our earlier single sensor measurements (Tajmar et al, 2006) which was based on evaluating the maximum peaks versus the applied acceleration, which were overpredicted most probably due to oscillation of the sensor's tilting plates.

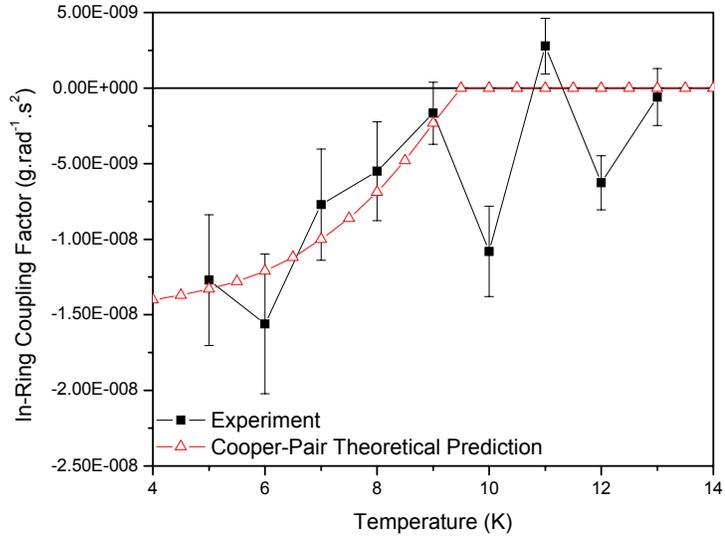

**FIGURE 9.** Variation of In-Ring Coupling Factor with Temp-erature and Cooper-Pair Theoretical Prediction according to Equ. (2) Multiplied by 2.

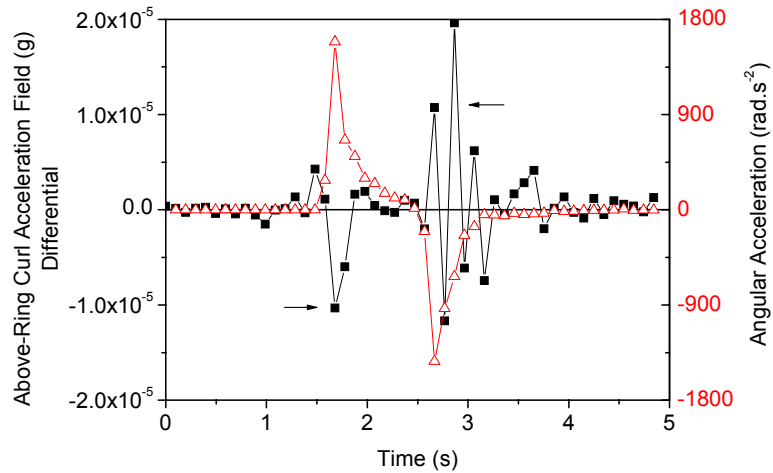

**FIGURE 10.** Signal Averaged over 24 Measurements at T=(4.5,6.5) K; Above-Ring Tangential All (Differential) Sensor Data (■) Versus Applied Angular Acceleration (Δ).

Fig. 10 shows a similar signal averaged analysis for the above-ring tangential sensor (also differential). In this graph all 4 tangential sensors were used, i.e. Sensor (1-3)+(4-2) (see Fig. 4) to measure a curl field all around the circle. The coupling factor while superconducting is $-6.06 \pm 1 \times 10^{-9}$ g.rad$^{-1}$.s$^2$. This is smaller than our result for the in-ring sensor, but this was expected from field expansion (the Z-component above the ring is much smaller than the in-ring value) and Equ. (2) with the different radial distance.

By modeling the gravitomagnetic field of the rotating superconductor similar to the one from a homogenously magnetized cylinder, we would have expected an above-ring value which is 0.3 times the in-ring value. Indeed, comparing the coupling factors from Fig. 8a and 10, the ratio is 0.27 which is very close to the estimate based on field expansion. The first tangential sensor peak in Fig. 10 seems much better defined than the second one which appears to be oscillating. That is because the observed signals here are smaller then the in-ring values and the noise contributions especially from the air motor are more dominant in the second peak. Also here a better sensor resolution would definitely help to improve the signal quality.

# GYROSCOPE MEASUREMENTS

The laser gyro should lead to very clean results as it is highly insensitive to background vibrations, our number one problem area with the accelerometers, and it's very accurate. They were implemented recently into our setup and the results presented in this section should be considered preliminary as only a few runs were carried out up to now. Nevertheless, they provide important information that also characterizes the mechanical environment that was present in the accelerometer measurements. The major difference for the gyro characterization is the fact that the gyros should measure a field directly proportional to the applied angular velocity of the superconductor whereas the accelerometer measurements were sensitive to the applied angular acceleration. In principle, this allows to average the signal over a much longer time (speed can be maintained constant but acceleration has to stop when maximum speed is reached), which can effectively reduce the noise.

Using again the analogy of the magnetized cylinder, we expected to see 50% of the gravitomagnetic in-ring value in Equ. (1) at the above-ring gyro and less than 3% on the reference position. Fig. 11 shows the above-ring and reference gyro versus the applied angular velocity for the same Niobium superconductor as used for the accelerometer measurements in a temperature range of 4.5 – 10 K. Overlaid is the Cooper-pair density prediction of Equ. (1) divided by a factor of four which gives a reasonable fit. The temperature dependence was modeled using the usual $(1-(T/T_c)^4)$ term. As we expected to see 50% of Equ. (1), the data underpredicts our theory by a factor of about 2 compared to the overprediction of a factor of 2 in the accelerometer measurements. We can see two hills/valleys on each profile. When the superconductor starts to rotate, the gyro measures a field dragging it with the rotating superconductor. As the ring is spinning up, the temperature is rising and the Cooper-pair density is decreasing accordingly. From the magnetic field sensor and the field coil, we know that the superconductor passed $T_c$ at 19.3 s which is exactly where the gyro signal goes down in the second profile. Immediately when the ring is slowed down, the temperature goes down again and the gyro measures a signal.

The fact that the above-ring laser gyro indeed seems to follow this Cooper-pair density is very encouraging. Especially it can be clearly seen that the reference gyro does not show a similar signal within a noise level of $3\times10^{-5}$ rad.s$^{-1}$. As both gyros are mounted on the same mechanical structure, it is impossible that the signal of the above-ring gyro is generated by a mechanical torque on the sensor chamber. The only possibility is that it sees a field emitted from the superconductor. The only field we know that would have such an effect on the gyro is a gravitomagnetic field (that is how it is measured on Gravity Probe-B). If this is correct, then our measurement shows that spacetime is dragged by the rotating superconductor. The reference gyro measurement rules out any mechanical acceleration larger than 1 µg at the sampling rate of 10 Hz, which is more than an order of magnitude below the observed effects from the accelerometers.

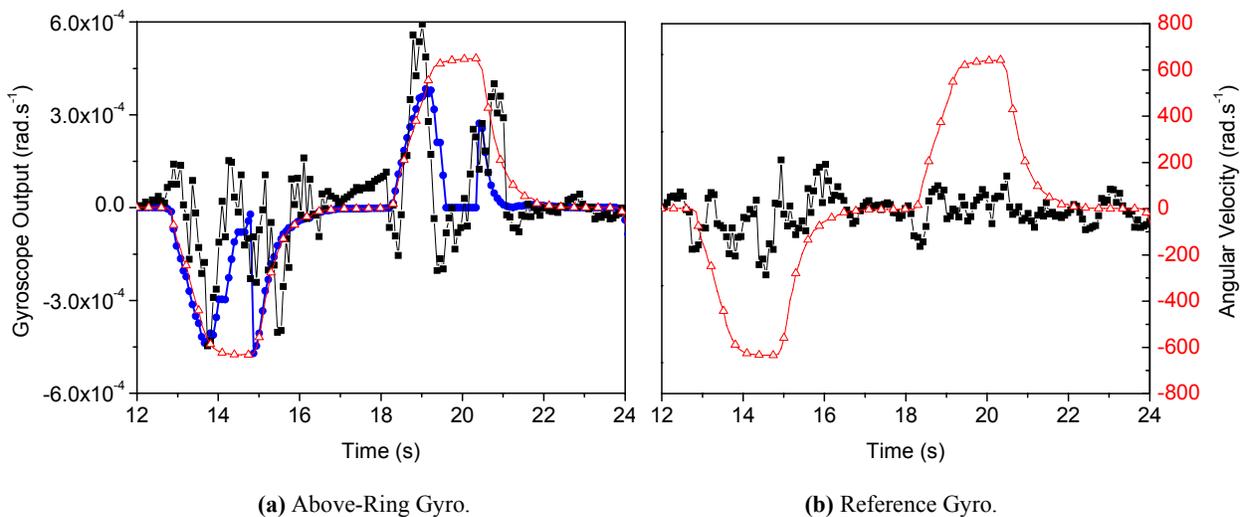

**(a)** Above-Ring Gyro.  **(b)** Reference Gyro.

**FIGURE 11.** Laser Gyro Data Sampled at 15 Hz with a 5 Point Moving Average Filter (■) Versus Applied Angular Velocity (Δ) and Compared to the Theoretical Prediction (●) of Equ. (1) Divided by Factor of 4 at T=(4.5,10) K.

The signal-to-noise level from this gyro data is better than 16:1. Notice that the sign of the angular velocity measured by the gyro is the same as the sign of the applied angular velocity to the superconductor, which is as expected from Equ. (1). This fits then also with the negative sign of the accelerometer measurements where the sign change is coming from the induction law.

Due to the importance of these conclusions, more measurements will be carried out in the near future to consolidate the gyro measurements. For comparison, Fig. 12a shows the above-ring gyro during the same run as with the Niobium superconductor in Fig. 11a but above the critical temperature. No signal change can be identified as expected. Another test run using the YBCO superconductor did not show any difference between above-ring and reference gyro down to 10 K (see Fig. 11b), well below YBCO's critical temperature. This was again expected as the ratio of $\rho^*/\rho$ is more than an order of magnitude below the one of Niobium and hence within the noise of the gyro.

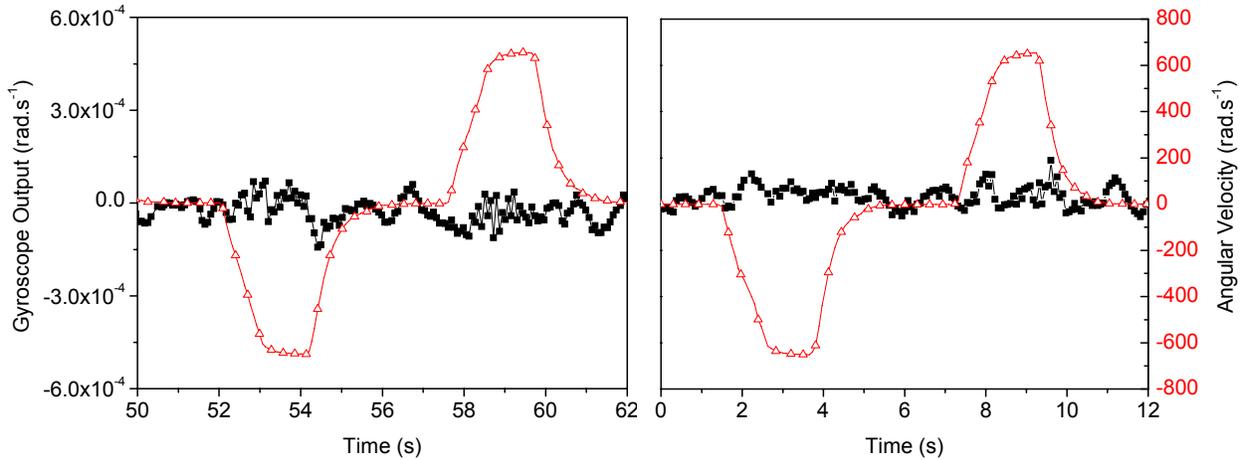

**(a)** Above-Ring Gyro with Nb at $T>T_c$; T=(18,20) K.   **(b)** Above-Ring Gyro with YBCO at $T<T_c$; T=(10,17) K.

**FIGURE 12.** Laser Gyro Data Sampled at 15 Hz with a 5 Point Moving Average Filter (■) Versus Applied Angular Velocity (Δ).

## DISCUSSION AND CONCLUSION OF RESULTS

We observed signals on tangential accelerometers mounted inside and above a superconducting Niobium ring when the ring is angularly accelerated. This signal is proportional to the applied acceleration and shows in the opposite direction. The signals got clearer when the accelerometers were operated in a curl-configuration where every sensor has a mirror partner to cut out mechanical offsets such as tilts. This showed that the acceleration we observe is a closed-loop field. On the other hand, preliminary measurements with a laser gyro mounted closely to a Niobium superconductor ring show a signal proportional to the applied angular velocity of the ring. Both accelerometer and gyro measurements seem to follow the Cooper-pair density of the superconductor. Sensors mounted on the same mechanical structure and further away than the sensors close to the superconductor do not show any signal within their resolution. Measurements with YBCO under similar conditions did not show these effects within our resolution.

What is the cause of these effects?

– Mechanical Artifacts: There is a connection between the rotating superconductor and the sensor chamber through the helium gas. Although the gas does not interact with the sensors as they are inside an evacuated chamber, it could drag the outer chamber walls. In order to characterize such artifacts, both accelerometer and gyro measurements were done with sensors also on the reference position to monitor the mechanical environment. Every mechanical rotation or tilt would be seen also on the reference sensor and the differential accelerometer measurement would show no signal. The fact that the reference gyro showed no signal at all

- within its resolution is reliable information that there is no mechanical movement of the sensor chamber, which is mounted with three solid steel bars to the ceiling for its stability.

- Electromagnetic Artifacts: Accelerometer offsets were characterized by applying magnetic fields. It was found out that a field of at least 100 mT would be required to induce an offset that has the same value as the signals we measured. A Hall sensor close to the accelerometers showed that during the air motor measurements, the magnetic field change during all runs was below the sensor's resolution of 1 µT. The influence on the gyros is even several orders of magnitude below. Electric field influence can be neglected as the sensors are mounted inside a grounded vacuum chamber acting as a Faraday cage.

- Helium Environment: The temperature range below 10 K also coincides with the range where liquid helium is strongly evaporated during the rotation of the superconductor, which creates an additional acoustic noise background in comparison to the runs at higher temperatures where only little liquid Helium evaporates. We know already that this acoustic noise, which causes small vibrations as it travels throughout the facility, can create DC accelerometer offsets in the hundred of µg range. If these offsets vary only by a few percent from sensor to sensor, it might still be possible that our results do not show the presence of the gravitomagnetic London moment but sensor induced offsets. However, this is very unlikely as this noise background does not depend on the sense of rotation, the observed effect on the other hand does. No vibration effect was found on the gyros. Moreover, as the YBCO measurements showed no effect, it demonstrates that the effect relates to the superconductor used and not to the helium environment.

From all analysis performed up to now, the most probable explanation of our results is the existence of the gravitomagnetic London moment. This is even further justified by the fact that the results seem to fit within a factor of 2 to our theoretical predictions in Equs. (1) and (2). The discrepancy between accelerometer and gyro measurements have to be further evaluated, probably higher resolution sensors will close the gap between them. Moreover, sensors at different positions measure different field strengths which closely resembles estimates based on standard field expansion.

It is very important that our results are investigated and verified in other laboratories to see if the effect can be reproduced in other facilities and probably different sensors to finally rule out any setup/facility induced effect. If our results hold true, then a rotating superconductor drags spacetime with it and it can be used to generate acceleration fields. That would obviously be of tremendous technological interest with the possibility to build a zero-g simulator on Earth or to move any object at a distance without physical contact to name a few examples. And our experimental results should not be too surprising as they are predicted by Einstein's general relativity theory and the presently observed amount of dark energy in the universe.

## NOMENCLATURE

$B_g$ = gravitomagnetic field (rad.s$^{-1}$)
g = gravitational field (in unit of Earth standard acceleration = 9.81 m.s$^{-2}$)
$\Lambda$ = cosmological constant (m$^{-2}$)
$\omega$ = Superconductor angular velocity (rad.s$^{-1}$)
$\dot\omega$ = Superconductor angular acceleration (rad.s$^{-2}$)
$\rho^*$ = Cooper-pair mass density (kg.m$^{-3}$)
$\rho$ = bulk density (kg.m$^{-3}$)
r = radial distance of sensor from center axis (m)
T = temperature of superconductor (K)
$T_c$ = critical temperature (K)

## ACKNOWLEDGMENTS

Part of this research was sponsored by the European Space Agency under GSP Contract 17890/03/F/KE and by the Air Force Office of Scientific Research, Air Force Material Command, USAF, under grant number FA8655-03-1-3075. The U.S. Government is authorized to reproduce and distribute reprints for Governmental purposes notwithstanding any copyright notation thereon. We would like to thank C.J. de Matos for many stimulating discussions and for starting the theoretical investigation together as well as I. Vasiljevich for helpful inputs and critical analysis. Also the efforts from T. Sumrall, M. Fajardo and R. Sierakowski to support our experimental efforts are greatly appreciated.

# REFERENCES

Argyris, J., Ciubotariu, C., "Massive Gravitons in General Relativity," *Aust. J. Phys.*, **50**, 879-891, (1997).

Becker, R., Heller, G., and Sauter, F., "Über die Stromverteilung in einer Supraleitenden Kugel, " *Z. Physik*, **85**, 772-787, (1933).

de Matos, C.J., and Tajmar, M., "Gravitomagnetic London Moment and the Graviton Mass inside a Superconductor," *Physica C*, **432**, 167-172, (2005).

de Matos, C.J., "Linearization of Einstein Field Equations with a Cosmological Constant in a Flat Background," (2006), http://arxiv.org/abs/gr-qc/0609116, accessed October 17, 2006.

Forward, R.L., "General Relativity for the Experimentalist," *Proceedings of the IRE*, 892-586, (1961).

Mashhoon, B., "On the Gravitational Analogue of Larmor's Theorem, " *Phys. Lett. A*, **173**, 347-354 (1993).

Spergel, D.N., et al., "Wilkinson Microwave Anisotropy Probe (WMAP) Three Year Results: Implications for Cosmology, " *Astrophy. J. Suppl.*, **148**, 175 (2003).

Tajmar, M., and de Matos, C.J., "Coupling of Electromagnetism and Gravitation in the Weak Field Approximation," *Journal of Theoretics*, **3**(1), (2001).

Tajmar, M., and de Matos, C.J., "Gravitomagnetic Field of a Rotating Superconductor and of a Rotating Superfluid," *Physica C*, **385**(4), 551-554, (2003).

Tajmar, M., and de Matos, C.J., "Extended Analysis of Gravitomagnetic Fields in Rotating Superconductors and Superfluids", *Physica C*, **420**(1-2), 56-60, (2005).

Tajmar, M., and de Matos, C.J., "Local Photon and Graviton Mass and its Consequences," (2006), http://arxiv.org/abs/gr-qc/0603032, accessed October 17, 2006.

Tajmar, M., and de Matos, C.J., "Gravitomagnetic Fields in Rotating Superconductors to Solve Tate's Cooper Pair Mass Anomaly," in the proceedings of *Space Technology and Applications International Forum (STAIF-2006)*, edited by M.S. El-Genk, AIP Conference Proceedings 813, Issue 1, Melville, New York, 2006, pp. 1415-1420.

Tajmar, M., "A note on the Local cosmological constant and the dark energy coincidence problem, " *Class. Quant. Grav.*, **23**, 1-5, (2006).

Tajmar, M., Plesescu, F., Marhold, K., and de Matos, C.J., "Experimental Detection of the Gravitomagnetic London Moment," (2006), http://arxiv.org/abs/gr-qc/0603033, accessed October 17, 2006.